\documentclass[sigconf]{acmart} 

\usepackage{amsmath}
\usepackage[ruled,vlined]{algorithm2e}
\usepackage{multirow}
\usepackage{subcaption}
\usepackage{kotex}
\usepackage{mathtools}
\DeclarePairedDelimiter{\abs}{\lvert}{\rvert}
\usepackage{enumitem}
\setlist[itemize]{leftmargin=3mm}
\usepackage[skip=2pt]{caption} 

\AtBeginDocument{%
  \providecommand\BibTeX{{%
    \normalfont B\kern-0.5em{\scshape i\kern-0.25em b}\kern-0.8em\TeX}}}

\copyrightyear{2020}
\acmYear{2020}
\setcopyright{iw3c2w3}
\acmConference[WWW '20]{Proceedings of The Web Conference 2020}{April 20--24, 2020}{Taipei, Taiwan}
\acmBooktitle{Proceedings of The Web Conference 2020 (WWW '20), April 20--24, 2020, Taipei, Taiwan}
\acmPrice{}
\acmDOI{10.1145/3366423.3380042}
\acmISBN{978-1-4503-7023-3/20/04}
\begin{document}

\title{Deep Rating Elicitation for New Users in Collaborative Filtering}

\newcommand{\rs}{RS\xspace}
\newcommand{\proposed}{DRE\xspace}
\author{Wonbin Kweon, SeongKu Kang, Junyoung Hwang, Hwanjo Yu\*}
\affiliation{%
   \institution{Pohang University of Science and Technology, South Korea}
   \{kwb4453, seongku, jyhwang, hwanjoyu\}@postech.ac.kr
}
\authornotemark[0]
\authornote{Corresponding Author}


\begin{abstract}
Recent recommender systems started to use \textit{rating elicitation}, which asks new users to rate a small seed itemset for inferring their preferences, to improve the quality of initial recommendations.
The key challenge of the rating elicitation is to choose the seed items which can best infer the new users’ preference.
This paper proposes a novel end-to-end \textbf{D}eep learning framework for \textbf{R}ating \textbf{E}licitation (\proposed), that chooses all the seed items at a time with consideration of the non-linear interactions.
To this end, it first defines categorical distributions to sample seed items from the entire itemset, then it trains both the categorical distributions and a neural reconstruction network to infer users’ preferences on the remaining items from CF information of the sampled seed items.
Through the end-to-end training, the categorical distributions are learned to select the most representative seed items while reflecting the complex non-linear interactions.
Experimental results show that \proposed outperforms the state-of-the-art approaches in the recommendation quality by accurately inferring the new users’ preferences and its seed itemset better represents the latent space than the seed itemset obtained by the other methods.
\end{abstract}



\maketitle

\section{Introduction}
Making accurate initial recommendations for newly joined users (so-called \textit{cold-start users}) has remained a long-standing problem in recommender systems (\rs) \cite{coldstart02}.
To infer the preferences of new users who have not interacted with any item yet, most methods have resorted to user side information such as demographic profiles (e.g., age, gender, and profession) \cite{cold_user_sideinfo1} and social relationships \cite{cold_social1, cold_social2}.
Nowadays, however, collecting personal information is challenging due to privacy issues, and a large number of users are reluctant to provide their private information to the website \cite{privacy02}. 

\begin{figure}[t]
  \includegraphics[width=0.46\textwidth]{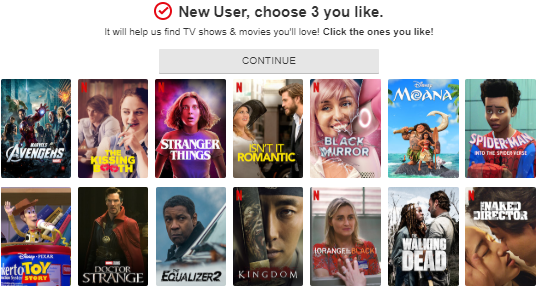}
  \caption{Rating elicitation of Netflix. When a new user signs up, Netflix shows 78 seed items to the new user.\protect\footnotemark}
  \label{fig:netflix}
\end{figure}
\footnotetext{It is captured on Oct. 10, 2019}

To overcome this limitation, a few recent work have focused on \textit{rating elicitation} \cite{rashid2008learning, rbmf11, maxvol16, iui02}, which asks new users to rate a small \textit{seed itemset}, to infer their preferences without using any personal information.
In Figure \ref{fig:netflix}, for example, Netflix initially shows some seed items to newly joined users and asks them to choose their favorite items.
Afterward, Netflix provides personalized recommendations based on the chosen items by each user.
Since \rs should keep the size of seed itemset small not to bother new users, the key challenge is to find small seed itemset that can provide enough information to infer new users' preferences.

Early methods for selecting the seed itemset \cite{iui02, rashid2008learning} have focused on the statistic-based scoring strategy that measures how representative each item is and selects top-$k$ items as the seed set.
They design score function based on items' popularity or variance (or entropy) of ratings.
However, those methods do not consider the interactions between the seed items, which might result in redundant selection and further poor recommendation accuracy.

To tackle this challenge, several recent work \cite{rbmf11, maxvol16} focus on the maximal-volume approach which finds the most representative combination of item latent vectors.
Specifically, they first define the amount of the representativeness as the volume of the parallelepiped spanned by the linear combination of selected latent vectors.
Then, they adopt a greedy fashion that adds representative items into the set one by one.
After finding the seed itemset, they predict new users’ preferences for the remaining items by linearly aggregating their feedback on the seed items.
This approach achieves state-of-the-art recommendation performance on the rating elicitation task.

Despite their effectiveness, we argue that the existing methods are not sufficient to yield satisfactory performance for the following reasons.
First, they only consider the linear interactions between the items both for finding the seed itemset and inferring the new users' preferences, which may not be sufficient to capture the complex structure of the collaborative filtering (CF) information (i.e., the user-item interactions).
Second, they adopt the greedy approach that makes use of the best short-term choice at each step, which cannot take into account the interactions of the whole seed items at a time, and may further degrade the quality of recommendation.

In this paper, we propose a novel end-to-end \textbf{D}eep learning framework for \textbf{R}ating \textbf{E}licitation (\proposed) that chooses all the seed items at a time with consideration of the non-linear interactions between the items. 
To this end, \proposed first defines the same number of categorical distributions as the number of seed items, and samples each seed item from each distribution.
\proposed then trains both the categorical distributions and a neural reconstruction network to infer the users’ preferences on the remaining items from CF information of the sampled seed items.
Through the end-to-end training, the categorical distributions are learned to select the most representative set of items while reflecting the non-linear interactions of the seed items.
Lastly, \proposed provides personalized initial recommendations for new users by using the trained reconstruction network and their feedback on the seed itemset.

However, sampling the seed items is a non-differentiable operation which would block the gradient flow and disable the end-to-end training.
\proposed uses a continuous relaxation of discrete distribution, Gumbel-Softmax \cite{gumbel16} whose parameter gradients can be computed via the reparameterization trick and trained by the backpropagation.
With the relaxation, \proposed successfully incorporates the non-differentiable operation into the network and takes the benefits of the end-to-end training.

Our extensive experiments demonstrate that DRE outperforms all other baselines. 
DRE finds the seed itemset that best infers the preference of new users, and provides more accurate initial recommendations based on their feedbacks than the state-of-the-art method does.
Furthermore, we qualitatively show that the seed itemset obtained by DRE is capable of better representing the latent space without the redundancy compared to the other methods.
We provide the source code of DRE for reproducibility.\footnote{\url{https://github.com/WonbinKweon/DRE_WWW2020}}

\section{Problem Formulation}
Let the set of users and items be denoted as $U=\{u_1, u_2,..., u_n\}$ and $V=\{v_1, v_2,...,v_m\}$ where $n$ and $m$ are the number of users and items, respectively. 
Let $R$ $\in \mathbb{R} ^{n \times m}$ denote the user-item rating matrix.
As we focus on implicit feedback, each element of $R$ has a binary value indicating whether a user has interacted with an item or not. 
Let $r^{(u)}$ $\in \mathbb{R}^{m}$ denote the user rating vector (i.e., $u$-th row of $R$).
We define \textit{seed itemset} as a set of items that we ask new users to rate when they sign up for the recommender system. The set of indices of the seed items is denoted as $S$ with $\abs{S} = k$. We also define \textit{candidate items} as all items except the seed items.

Rating elicitation can be divided into three steps: 
(1) Select the seed itemset $S$ with rating history of training users
(2) Elicit ratings $z^{(u)} \in \mathbb{R}^k$ on $S$ from a new user 
(3) Predict the ratings of the new user on the candidate items and provide initial recommendations. 
Thus, we are interested in finding the optimal seed itemset $S$ and a reconstruction function $f_{\theta}(\cdot)$ by solving below optimization problem.
\begin{equation}
\operatorname*{min}_{S,\theta} \frac{1}{2} \| R-f_{\theta}( R[:,S] ) \|^2_F.
\end{equation}
Note that $R[:,S] \in \mathbb{R}^{n \times k}$ is in the numpy indexing notation\footnote{\url{https://docs.scipy.org/doc/numpy-1.13.0/reference/arrays.indexing.html}}, which means we take a submatrix of $R$ by taking only $k$ columns with the indices in $S$. 
Thus, $R[:,S]$ is the rating matrix of the seed itemset $S$. 
Then the reconstruction function $f_{\theta}(\cdot)$ predicts the ratings of the candidate items by reconstructing $R$ from $R[:,S]$.
In this paper, we call the reconstruction function $f_{\theta}(\cdot)$ a \textit{decoder}.

\section{Preliminaries}
In this section, we introduce existing maximal-volume approaches for selecting the seed itemset (Section 3.1) and the Gumbel-Softmax which allows the sampling process of \proposed to be differentiable for the backpropagation (Section 3.2).

\subsection{Maximal-Volume Approaches}
The state-of-the-art maximal-volume approaches \cite{rbmf11, maxvol16} are proposed to find the seed itemset with consideration of the linear interactions between the seed items.
They find the seed itemset and the decoder by solving the following optimization problem
\begin{equation}
\begin{aligned}
\min_{S,X} \quad & \frac{1}{2} \| R-R[:,S]X \|^2_F, \\
\end{aligned}
\end{equation}
where $X \in \mathbb{R}^{k \times m}$ corresponds to the decoder.
They first reduce the dimensionality of column space of $R$ from $n$ to $k$ by using the rank-$k$ Singular Value Decomposition (SVD)
\begin{equation}
R = U V^T, \quad  \text{where } U \in \mathbb{R}^{n \times k}\text{ and } V^T \in \mathbb{R}^{k \times m}.
\end{equation}
Then they select $k$ representative columns (i.e., the items) of $V^T$ that can represent all the remaining columns. 
To this end, they adopt Maxvol algorithm \cite{maxvol10} which searches for the columns that maximize the volume of the parallelepiped spanned by the linear combination of them.
To maximize the volume, the columns should have large norms and should be evenly distributed to prevent linear dependency.
Maxvol algorithm adds representative items into $S$ one by one in a greedy fashion, because finding the globally optimal $S$ and $X$ is NP-hard \cite{maxvol16}.
After finding $S$, they compute the decoder $X$ with pseudo-inverse of $R[:,S]$ as $X = (R[:,S]^{T}R[:,S])^{-1}R[:,S]^{T}R$.
Lastly, they predict a rating vector of a new user based on the elicited feedback $z^{(u)}$ and the decoder $X$ by $\hat{r}^{(u)} = z^{(u)}X$.
By sorting the predicted ratings on the candidate items, they provide initial recommendations to the new user.

Overall, the maximal-volume approaches have two intrinsic flaws.
First, they cannot capture the non-linear interactions among the seed items, because both Maxvol algorithm and the decoder $X$ only consider the linear interactions of the column vectors. It may not be sufficient to capture the complex structure of the CF information. Second, they adopt the greedy search that incrementally increases the seed itemset, which cannot consider the interactions of the whole seed items at a time.
As a result, they may fall into the local optimum and fail to provide accurate recommendations.

\subsection{Gumbel-Softmax}
Gumbel-Softmax \cite{gumbel16} is a continuous distribution on the simplex that can approximate samples from a categorical distribution. 
If there is a categorical distribution with class probability $\pi=[\pi_1,\pi_2,...,\pi_m]$ for $m$ classes, we can express a categorical sample as a $m$-dimensional one-hot vector.
Gumbel-Max trick \cite{gumbelmax14} introduces a simple way to draw a sample $y$ from the categorical distribution with class probability $\pi$:
\begin{equation}
y = \textrm{one\_hot} \left( \operatorname*{arg\,max}_{j}[g_j+\textrm{log}\pi_{j}] \right),
\end{equation}
where $g_j$ is i.i.d drawn from Gumbel distribution with $\mu=0$, $\beta=1$\footnote{$g_j=-\text{log}(-\text{log}(u))$, where $u$ is sampled from $Uniform(0,1)$}. 
Gumbel-Softmax uses the softmax function as a continuous approximation of $ \operatorname*{arg\,max}$ operation, and get the approximated one-hot representation of the sample $y$:
\begin{equation}
y_j = \frac{\textrm{exp}((\textrm{log}\pi_j+g_j)/\tau)}{\Sigma_{l=1}^m \textrm{exp}((\textrm{log}\pi_l+g_l)/\tau)} \quad \textrm{for } j=1,...,m,
\end{equation}
where $\tau$ is a hyper-parameter for the temperature of the softmax. 
If we take very small $\tau$, $y \in \mathbb{R}^{m}$ will become one-hot and Gumbel-Softmax distribution becomes identical to the categorical distribution.
In this paper, we want to sample $k$ seed items from $k$ different categorical distributions. 
Therefore, we extend $\pi$ and $y$ to the 2-dimensional matrix from the 1-dimensional vector. 
Then we have $\Pi \in \mathbb{R}^{k \times m}$ and $Y \in \mathbb{R}^{k \times m}$. 
Each row of $\Pi$ represents the class probability of a categorical distribution and each row of $Y$ represents the approximated one-hot representation from each categorical distribution.

\section{Method}
Our proposed end-to-end \textbf{D}eep learning framework for \textbf{R}ating \textbf{E}licitation (\proposed) chooses all the seed items at a time while considering non-linear interactions of the items. 
\proposed consists of two modules: an encoder and a decoder, as illustrated in Figure 2. 
The encoder is a module for selecting the seed itemset and the decoder is a module for predicting the ratings for the candidate items by reconstructing the user rating vector $r^{(u)}$ from the feedback on the seed itemset $z^{(u)}$.

\begin{figure}[t]
  \centering
  \includegraphics[width=0.45\textwidth]{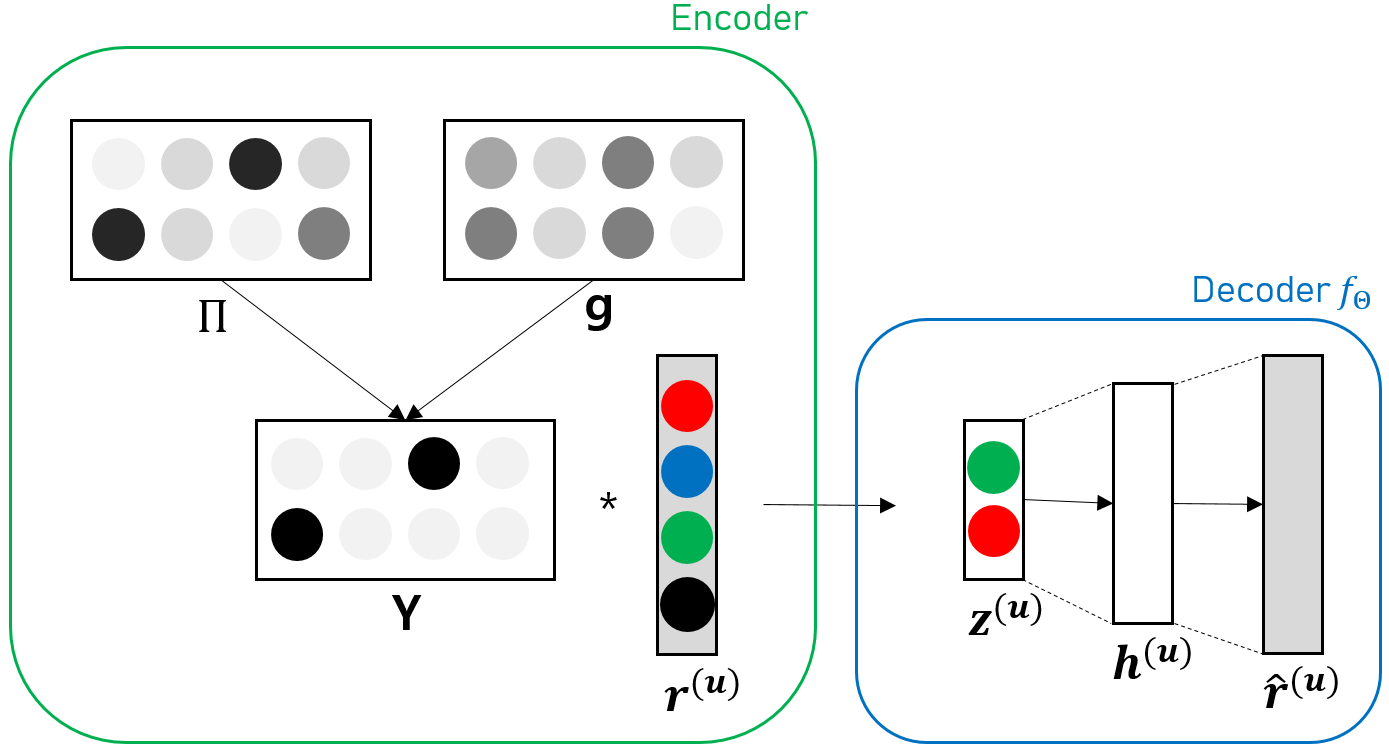}
  \caption{The architecture of DRE. $r^{(u)}$ is a user rating vector, $\Pi$ and $Y$ are introduced in Section 3.2. In the Figure, $Y \in \mathbb{R}^{2 \times 4}$ is a concatenation of two one-hot vectors $[0,0,1,0]$ and $[1,0,0,0]$ from Gumbel-Softmax of $\Pi$ and $g$. After multiplying $Y$ and $r^{(u)}$, $z^{(u)}$ has the third and the first values of $r^{(u)}$, which are the green circle and the red circle.}
\end{figure}

\subsection{Encoder}
The encoder samples $k$ seed items with a matrix $\Pi \in \mathbb{R}^{k \times m}$ which consists of $k$ trainable categorical distributions.
Each row $\Pi_i$ represents the class probability of a $m$-dimensional categorical distribution. 
We sample one item from each row $\Pi_i$, so that we draw $k$ items in total. 
However, the sampling process is non-differentiable, which would block the gradient flow and disable the end-to-end training.
We adopt a continuous relaxation of the discrete distribution by using Gumbel-Softmax.
With the relaxation, the categorical distributions $\Pi$ can be trained by the backpropagation.

We get $Y \in \mathbb{R}^{k \times m}$ from $\Pi$ and $g$ by using Gumbel-Softmax as
\begin{equation}
Y_{ij} = \frac{\textrm{exp}((\textrm{log}\Pi_{ij}+g_{ij})/\tau)}{\Sigma_{l=1}^m \textrm{exp}((\textrm{log}\Pi_{il}+g_{il})/\tau)} \quad \textrm{for i}=1,...,k, \textrm{ j}=1,...,m.
\end{equation}
\noindent With small $\tau$, $Y_i \in \mathbb{R}^{m}$ is an approximation of an $m$-dimensional one-hot vector sampled from the categorical distribution with class probability $\Pi_i$. 
In the same manner, $Y \in \mathbb{R}^{k \times m}$ consists of $k$ approximated one-hot vectors. 
Then we get $z^{(u)} \in \mathbb{R}^k$ by multiplying $Y$ and $r^{(u)}$
\begin{equation}
z^{(u)} = Y \cdot r^{(u)}.
\end{equation}
Note that $z^{(u)}$ has the $k$ elements of $r^{(u)}$, because each row of $Y$ is a one-hot vector.
Therefore, we can treat $z^{(u)}$ as ratings on the seed items.

\subsection{Decoder}
The decoder $f_{\theta}(\cdot)$ reconstructs user rating vector $r^{(u)}$ from the feedback $z^{(u)}$ by using the non-linear interactions between the seed items. 
To this end, we use the fully-connected neural network that is widely used in autoencoder-based CF methods to reconstruct user rating vectors \cite{autorec15, cdae16, CVAE17, vae18}.
 
In this paper, we reconstruct $r^{(u)}$ with a 2-layer fully-connected network
\begin{equation}
\begin{split}
  h^{(u)} = \sigma(W_1^T z^{(u)} + b_1) \in \mathbb{R}^{d}\\
  \hat{r}^{(u)} = \sigma(W_2^T h^{(u)} +b_2) \in \mathbb{R}^{m},
\end{split}
\end{equation}
where $\sigma$ is the sigmoid function, $W_1 \in \mathbb{R}^{k \times d}$ and $W_2 \in \mathbb{R}^{d \times m}$ are weight matrices, $b_1 \in \mathbb{R}^{d}$ and $b_2 \in \mathbb{R}^{m}$ are biases for the fully-connected layers, $k$ is the size of the seed itemset, $d$ is the dimension of the hidden layer and $m$ is the number of all items.

 \subsection{Model Training}
After the decoder reconstructs the user rating vector, we compute Mean Squared Error (MSE) loss for the end-to-end training of \proposed.
\begin{equation}
\operatorname*{min}_{\Pi,\theta} \mathcal {L} = \frac{1}{|U_{T}|} \displaystyle\sum_{u \in U_{T}} \| r^{(u)} - \hat{r}^{(u)} \|^2_F,
\end{equation}
where $\Pi$ represents the categorical distributions of the encoder, $\theta$ denotes all parameters of the decoder, and $U_T$ is a set of training users.
Note that $\Pi$ and $\theta$ are trained together in the end-to-end manner with the backpropagation.
Through the end-to-end training, $\Pi$ is learned to select the most representative set of items while reflecting the non-linear interactions of the items.

Since we have to deal with thousands of items, we use simple exponential annealing for $\tau$ as $\tau=T_0(\frac{T_E}{T_0})^{\frac{e}{E}}$, where $e$ is the current epoch, $E$ is the number of training epochs, $T_0$ is initial temperature and $T_E$ is the last temperature $(T_0 >> T_E)$. With a large $\tau$ the model can explore the combinations of the whole itemset, with annealed small $\tau$ the model can select seed itemset. 

After the training is done, we re-train the decoder with the fixed encoder for small epochs.
Finally, we can extract the seed itemset from the trained $\Pi$. Since $\Pi_i$ represents the class probability of an $m$-dimensional categorical distribution, we can extract the final seed items by taking the index with the maximal probability from each distribution
\begin{equation}
\label{eq:seeditemset}
S_i = \operatorname*{arg\,max}_{j \in \{1,...,m\}} \Pi_{ij} \quad \textrm{for i}=1,...,k,
\end{equation}
where $S_i$ is the $i$-th seed item.

\subsection{Recommendation for New Users}
When a new user $u$ signs up for the recommender system, we ask the new user to rate the seed itemset $S$. 
After getting feedback $z^{(u)}$ on the seed itemset, we predict the ratings for the candidate items by using the trained decoder $f_\theta(\cdot)$
\begin{equation}
\hat{r}^{(u)} = f_{\theta}(z^{(u)}) \in \mathbb{R}^{m}.
\end{equation}
Finally, we provide a top-N recommendation by sorting $\hat{r}^{(u)}$ in a descending order.

\section{Experiments}
In this section, we present experimental results supporting that \proposed outperforms state-of-the-art approaches in various aspects.

\subsection{Datasets}
We use four real-world datasets of MovieLens 1M\footnote{https://grouplens.org/datasets/movielens/}, CiteULike \cite{CUL13}, Yelp \cite{yelp16} and MovieLens 20M. 
For the explicit feedback datasets, we convert the ratings over 3.5 to 1 and otherwise to 0 as done in \cite{cml17, bpr09}. 
Also, we only keep users who have at least five ratings for MovieLens 1M and CiteULike, twenty ratings for Yelp and MovieLens 20M. 
Data statistics after the preprocessing are presented in Table \ref{tbl:statistic}.
For each dataset, we split all users into the training set (80\%) and the test set (20\%).
We also take 10\% of the training users for validation. 

\subsection{Metrics}
As we focus on the top-N recommendation task based on implicit feedback, we evaluate the performance of each method by using two ranking metrics: Precision (P@N) \cite{pre16} and Normalized Discounted Cumulative Gain (NDCG@N) \cite{NDCG02}.
P@N measures how many items in top-N are actually interacted with the test user, NDCG@N assigns higher weights on the upper ranked items.
For each test user $u$, we get the ranked item list $\omega^{(u)}$ as described in Section 4.4. 
Formally, we define $\omega^{(u)}_n$ as the item at rank $n$, $\mathbb{I}[\cdot]$ as the indicator function, and $V^{(u)}$ as the set of candidate items that user $u$ interacted with (i.e., the ground-truth items). 
P@N for user $u$ is defined as
\begin{equation}
 \text{P@N}(\omega^{(u)}, V^{(u)}) = \frac{1}{N} \sum_{n=1}^N \mathbb{I}[\omega^{(u)}_n \in V^{(u)}].
\end{equation}
DCG@N for user $u$ is defined as
\begin{equation}
 \text{DCG@N}(\omega^{(u)}, V^{(u)}) = \displaystyle\sum_{n=1}^N \frac{\mathbb{I}[\omega^{(u)}_n \in V^{(u)}]}{log(n+1)}.
\end{equation}
NDCG@N is the normalized DCG@N after dividing by the best possible DCG@N, where all the ground-truth items are ranked at the top. 
We compute the above metrics for each test user, then report the average score.
We conduct experiments for N=10, 20, 50, 100 and report the results only for N=10, 20 due to the lack of space.
The improvements (\%) are similar across different Ns.

\begin{table}[t]
 \renewcommand{\arraystretch}{0.7}
  \caption{Data Statistics after the preprocessing}
  \begin{tabular}{ccccc}
    \toprule
    Dataset & \#Users & \#Items & \#Ratings & Sparsity \\
    \midrule
    MovieLens 1M & 6,028 & 3,533 & 575,242 & 97.30\% \\
    CiteULike & 6,315 & 25,385 & 136,268 & 99.91\% \\
    Yelp & 25,658 & 8,612 & 570,455 & 99.74\% \\
    MovieLens 20M & 99,220 & 20,660 & 9,478,902 & 99.54\% \\
    \bottomrule
  \end{tabular}
    \label{tbl:statistic}
\end{table}

\begin{table*}[t]
 \renewcommand{\arraystretch}{0.5}
  \caption{Experiment results with $k$=50. Results of \proposed and the best baseline are in bold face. \textit{Improv.} denotes the improvement of \proposed over the best baseline. $*$, $**$, and $***$ indicate $p \leq $ 0.05, $p \leq $ 0.01, and $p \leq $ 0.005 for the paired t-test of \proposed vs. the best baseline.}
  \begin{tabular}{c|c|ccccccc|cc}
    \toprule 
     Dataset & Metrics & MOSTPOP& RAN++ & POP++ & RBMF & RBMF++ & RMVA & RMVA++ & DRE & \textit{Improv.}  \\
    \midrule
    \bottomrule
    \multirow{4}{*}{MovieLens 1M} & P@10 & 0.3733 & 0.4267 & 0.4011 & 0.5047 & 0.5027 & \textbf{0.5063} & 0.5033 & \textbf{0.5396} & 6.56\%{***} \\
    & P@20 & 0.3331& 0.3880 & 0.3557 & \textbf{0.4373} & 0.4366 & 0.4363 & 0.4325 & \textbf{0.4734} & 8.25\%{***} \\
    & NDCG@10 & 0.3921& 0.4439 & 0.4235 & 0.5359 & 0.5365 & \textbf{0.5387} & 0.5357 & \textbf{0.5688} & 5.59\%{***} \\
    & NDCG@20 & 0.3629& 0.4146 & 0.3935 & 0.4871 & 0.4883 & \textbf{0.4887} & 0.4852 & \textbf{0.5197} & 6.34\%{***} \\
    \midrule
    \multirow{4}{*}{CiteULike} & P@10 & 0.0073 &0.0146& 0.0365 & 0.0465 & 0.0528 & 0.0498 & \textbf{0.0581} & \textbf{0.0701} & 20.69\%{***} \\
    & P@20 & 0.0049 &0.0136& 0.0304 & 0.0366 & 0.0392 & 0.0380 & \textbf{0.0410} & \textbf{0.0471} & 14.90\%{***} \\
    & NDCG@10 & 0.0121 &0.0159& 0.0431 & 0.0590 & 0.0649 & 0.0610 & \textbf{0.0691} & \textbf{0.0912} & 31.98\%{***} \\
    & NDCG@20 & 0.0130 & 0.0181&0.0430 & 0.0569 & 0.0611 & 0.0582 & \textbf{0.0652} & \textbf{0.0842} & 29.14\%{***} \\
    \midrule   
    \multirow{4}{*}{Yelp} & P@10 & 0.0424&0.0496 & 0.0478 & 0.0602 & 0.0591 & 0.0608 & \textbf{0.0609} & \textbf{0.0633} & 4.01\%{**} \\
    & P@20 & 0.0371&0.0429 & 0.0393 & 0.0499 & 0.0498 & 0.0503 & \textbf{0.0505} & \textbf{0.0524} & 3.76\%{*} \\
    & NDCG@10 &0.0461& 0.0542 & 0.0525 & 0.0692 & 0.0674 & \textbf{0.0694} & 0.0690 & \textbf{0.0705} & 1.56\% \\
    & NDCG@20 &0.0477& 0.0548 & 0.0531 & 0.0679 & 0.0673 & 0.0681 & \textbf{0.0684} & \textbf{0.0696} & 1.75\% \\
    \midrule    
    \multirow{4}{*}{MovieLens 20M} & P@10 & 0.3853&0.4007 & 0.4063 & 0.5064 & 0.5049 & \textbf{0.5116} & 0.5113 & \textbf{0.5712} & 11.65\%{***} \\
    & P@20 & 0.3276&0.3575 & 0.3585 & 0.4312 & 0.4316 & \textbf{0.4382} & \textbf{0.4382} & \textbf{0.4875} & 11.24\%{***} \\
    & NDCG@10 &0.4071& 0.4196 & 0.4253 & 0.5414 & 0.5401 & \textbf{0.5444} & 0.5434 & \textbf{0.6077} & 11.62\%{***} \\
    & NDCG@20 &0.3596& 0.3826 & 0.3899 & 0.4806 & 0.4806 & \textbf{0.4851} & 0.4846 & \textbf{0.5373} & 10.76\%{***} \\
    \bottomrule
  \end{tabular}
\end{table*}

\subsection{Baselines}
To show the superiority of the proposed model, we use three groups of baseline methods: non-elicitation method, statistic-based methods, and maximal-volume approaches. 
Note that the methods with `++' use the non-linear decoder that has the same structure of that of \proposed (i.e., the fully-connected network).

The first group is the non-elicitation method, which does not use rating elicitation and produces non-personalized initial recommendations.
\begin{itemize}
    \item \textbf{MOSTPOP}: Without the rating elicitation process, it produces a ranked list of candidate items by sorting their popularity (\#ratings). 
    The candidate items are the same as those of \proposed. 
\end{itemize}

The second group is the statistic-based methods, which do not consider the interactions between the seed items.
\begin{itemize}
    \item \textbf{RAN++ \cite{iui02}} : Randomly select $k$ seed items from the entire itemset. 
    \item \textbf{POP++ \cite{iui02}} : Select the most popular $k$  items as a seed itemset. 
\end{itemize}

The last group includes the state-of-the-art maximal-volume approaches.
These methods select the seed itemset in a greedy fashion with consideration of the linear interactions of items.
Also, we include two variants that use the non-linear decoder to verify the effectiveness of the end-to-end training.
\begin{itemize}
    \item \textbf{RBMF \cite{rbmf11}} : Representative Base Matrix Factorization method for rating elicitation, which is introduced in Section 3.1.
    \item \textbf{RBMF++} : A variant of RBMF. After finding the seed itemset, it uses the non-linear decoder.
    \item \textbf{RMVA \cite{maxvol16}} : Our main competitor which is the state-of-the-art method for rating elicitation.
    It alleviates the squareness of RBMF by introducing the rectangular matrix volume which is a generalization of the usual determinant.
    \item \textbf{RMVA++} : A variant of RMVA. After finding the seed itemset, it uses the non-linear decoder.
\end{itemize}

Note that we do not compare DRE with personalized rating elicitation methods, because they use users' rating history \cite{iui02, sys14} or side information \cite{side13}.
DRE is for new users without any rating history and side information. 

\subsection{Implementation Details}
We use PyTorch \cite{pytorch19} and Adam optimizer \cite{adam14} for the proposed model and all baselines. 
For each dataset, hyper-parameters are tuned by using grid searches on the validation set. 
The learning rate for the Adam optimizer is chosen from $\{$0.1, 0.05, 0.01, 0.005, 0.004, 0.003, 0.002, 0.001$\}$.
We use a 2-layer fully-connected network as the decoder of \proposed. 
It has the shape of $[k \rightarrow d \rightarrow m]$.
The dimension of the hidden layer $d$ is chosen from $\{$200, 300, 500$\}$. 
For the temperature annealing, $T_0$ is chosen from $\{1, 10, 20, 50\}$ and $T_E$ is chosen from $\{$0.01, 0.1, 0.5, 1, $T_0$$\}$.
We find the best epoch in $\{ 200, 400, 1000, 2000 \}$.
For RMVA, we tune the decomposition rank of SVD for every possible number (i.e., any smaller number than $k$).
Lastly, we report the average value of five iterations for all methods.

\begin{figure*}[t]
\begin{subfigure}[t]{0.5\linewidth}
    \includegraphics[width=\linewidth]{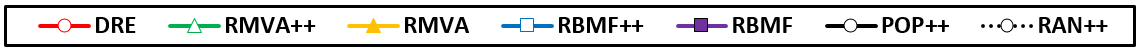}
\end{subfigure}

\begin{subfigure}[t]{0.241\linewidth}
    \includegraphics[width=\linewidth]{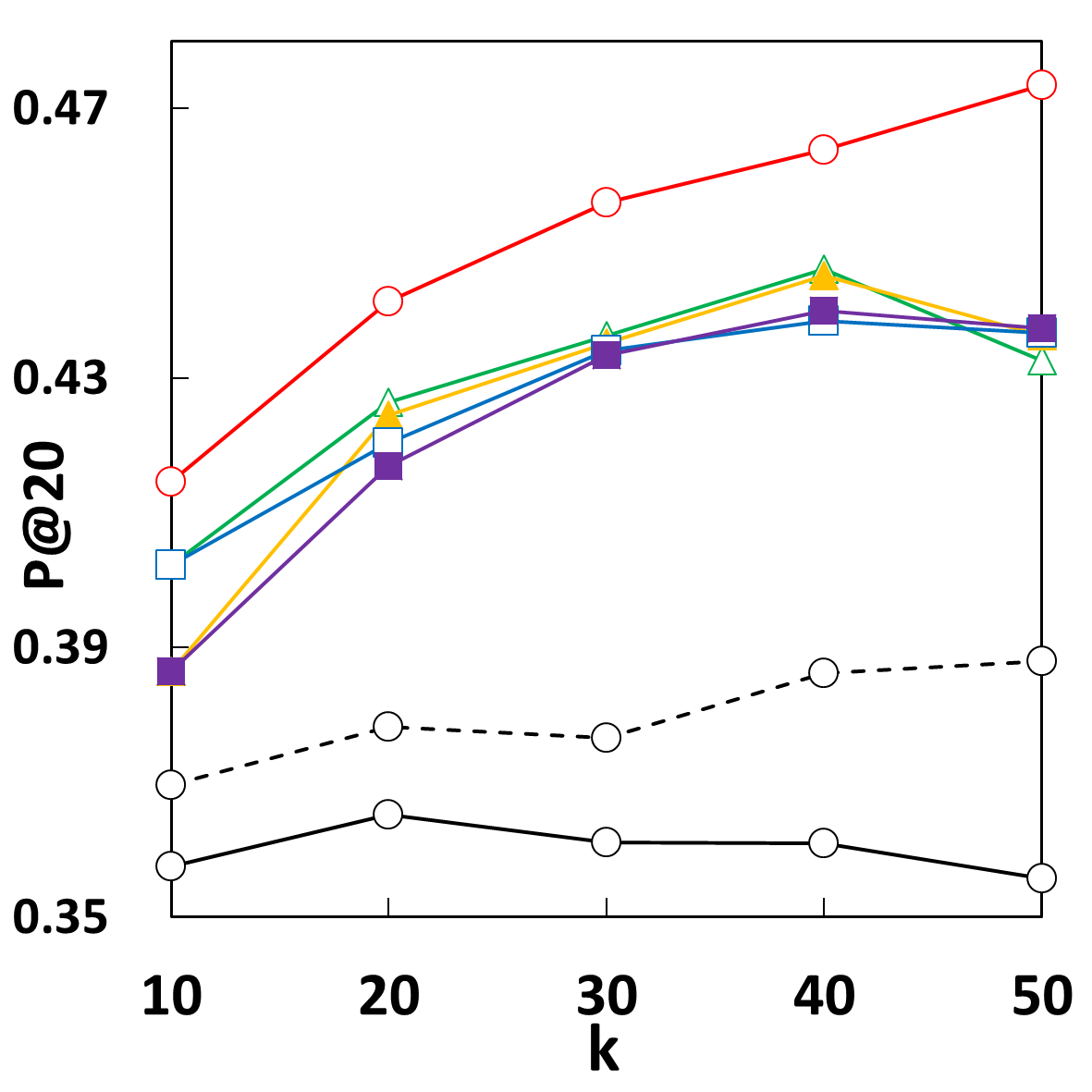}
    \caption{MovieLens 1M}
\end{subfigure}
\begin{subfigure}[t]{0.236\linewidth}
    \includegraphics[width=\linewidth]{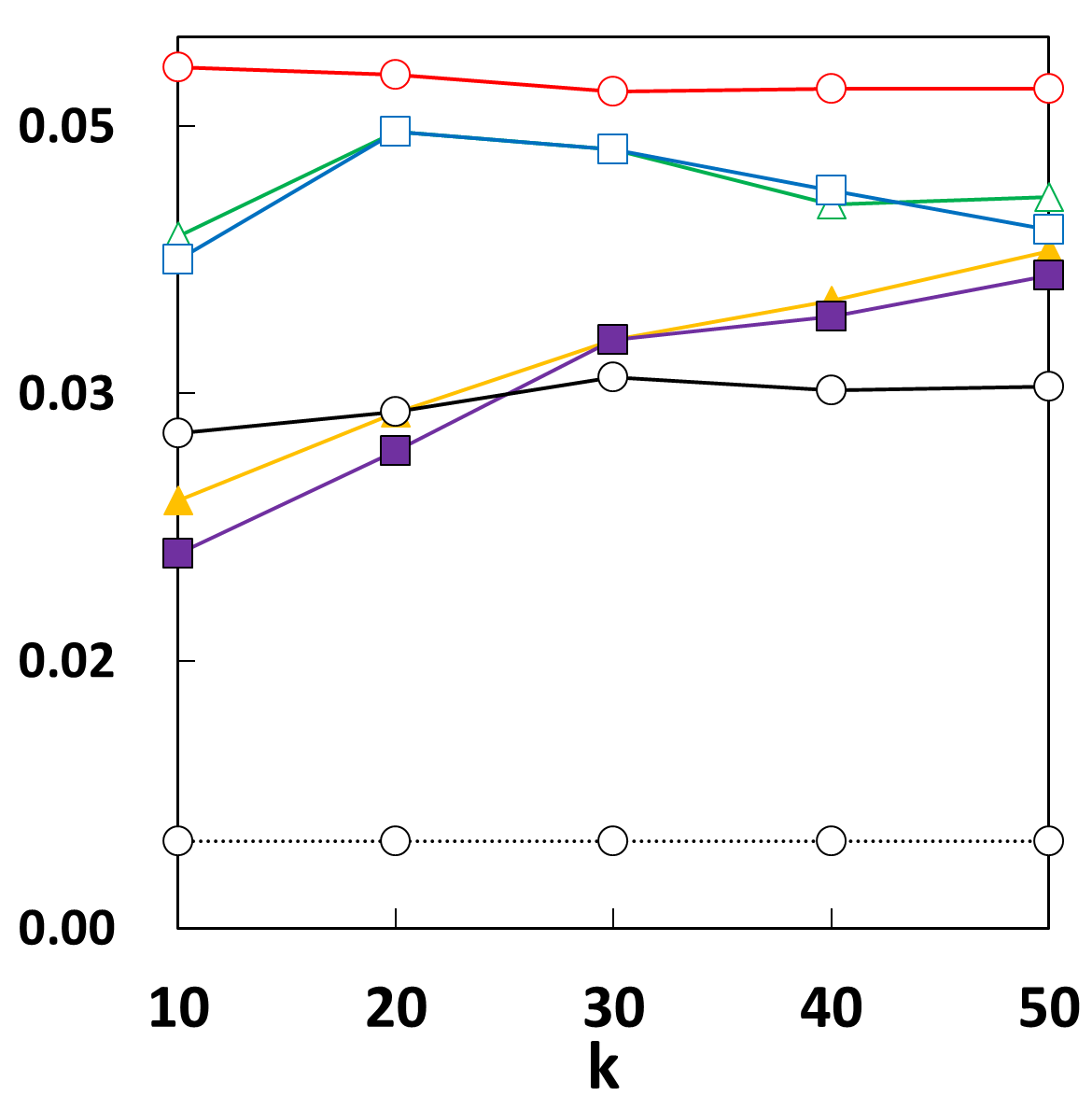}
    \caption{CiteULike}
\end{subfigure} 
\begin{subfigure}[t]{0.24\linewidth}
    \includegraphics[width=\linewidth]{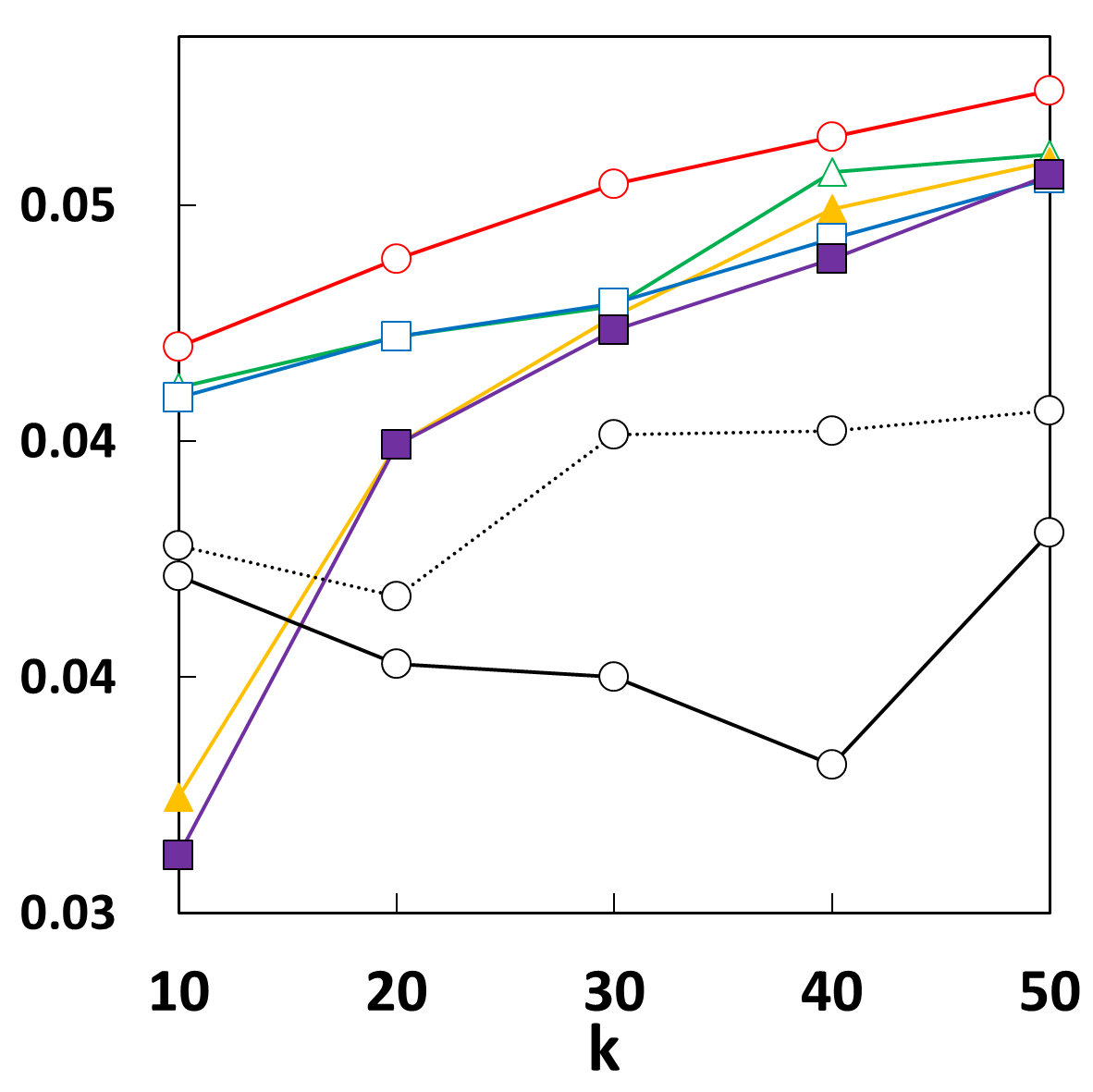}
    \caption{Yelp}
\end{subfigure} 
\begin{subfigure}[t]{0.243\linewidth}
    \includegraphics[width=\linewidth]{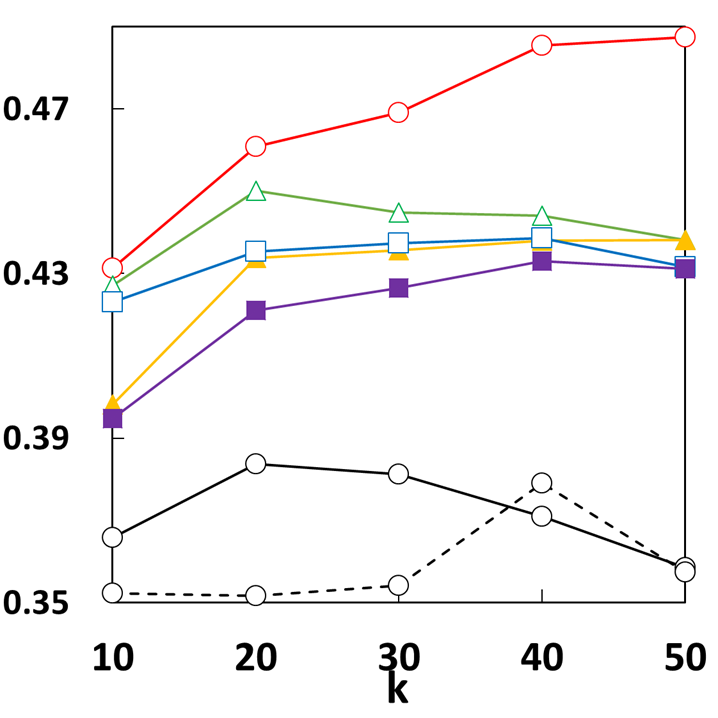}
    \caption{MovieLens 20M}
\end{subfigure} 
\caption{P@20 of DRE and baselines with different $k$ on four datasets.}
\end{figure*}

\begin{figure*}[t]
\begin{subfigure}[t]{0.33\linewidth}
    \includegraphics[width=\linewidth]{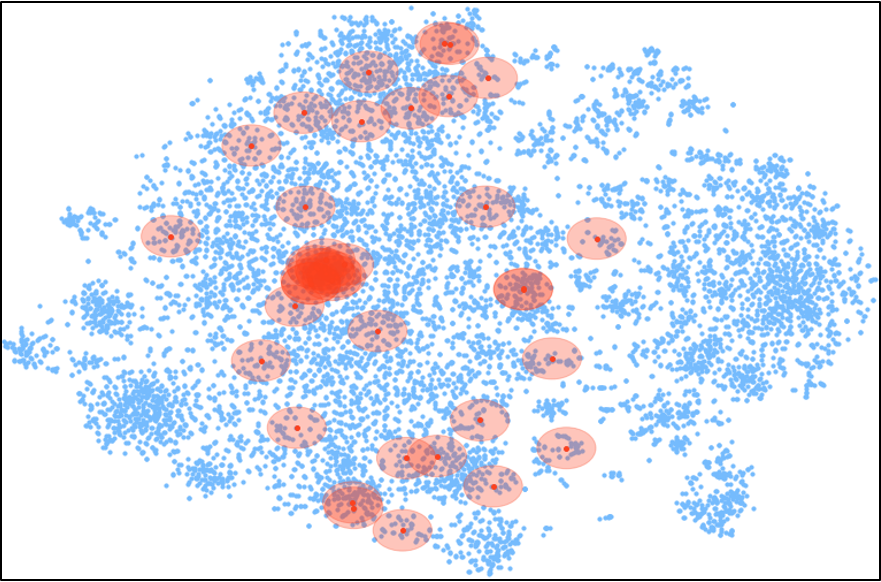}
    \caption{DRE}
\end{subfigure}%
\begin{subfigure}[t]{0.333\linewidth}
    \includegraphics[width=\linewidth]{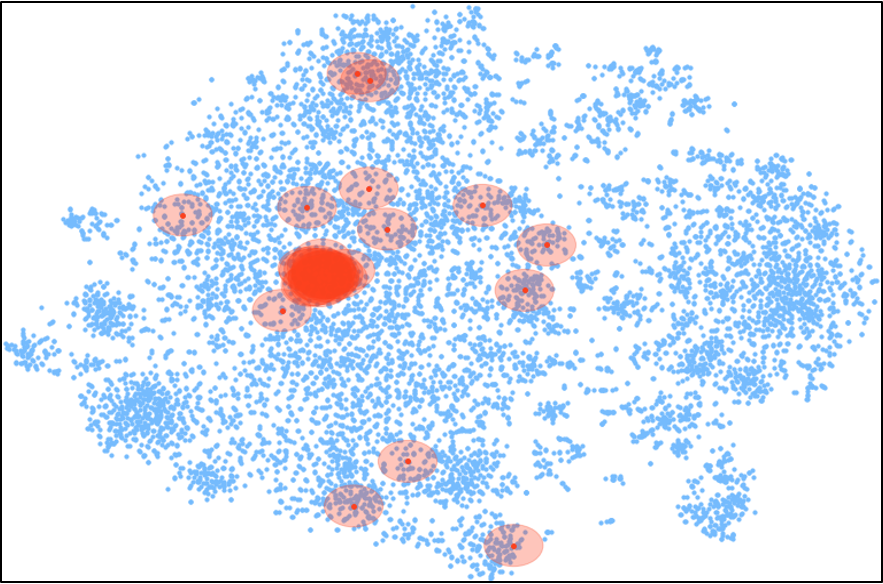}
    \caption{RMVA}
\end{subfigure} 
\begin{subfigure}[t]{0.333\linewidth}
    \includegraphics[width=\linewidth]{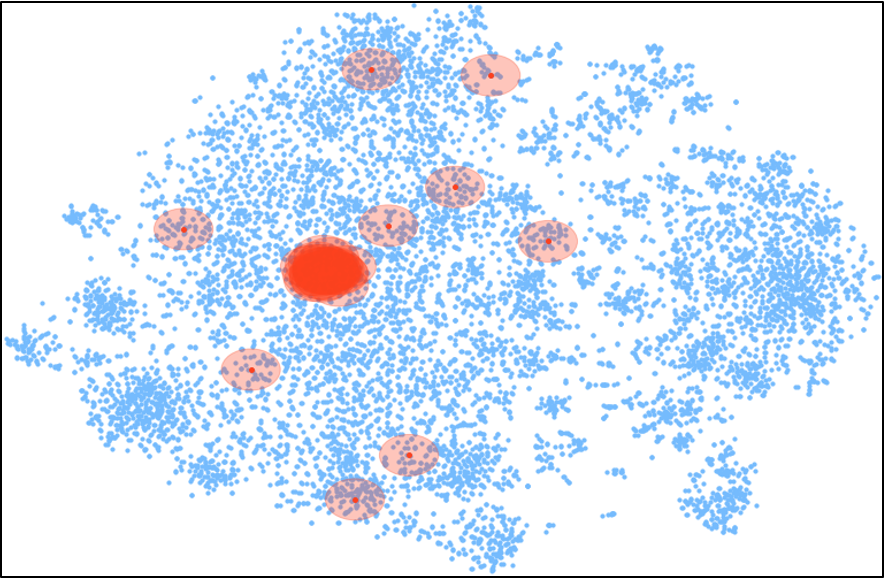}
    \caption{RBMF}
\end{subfigure} 
\caption{Location of the seed itemset on MovieLens 1M with $k$=40. Blue dots are latent vector of users and items, the red circles are the seed items.}
\end{figure*}
\subsection{Comparing with baselines}
We evaluate the top-N recommendation performance of \proposed and baseline methods in two ranking metrics. 
Table 2 shows the quantitative results and Figure 3 shows the result with the different sizes of the seed itemset $k$. 
In summary, the proposed model achieves the best results in all the metrics, especially on CiteULike dataset (by up to 31.98\% improvement for NDCG@10 compared to the best baseline).
Also, \proposed outperforms all the baselines with different $k$ in Figure 3. 
We analyze the results from various perspectives.

\vspace{1pt}
\noindent\textbf{Effectiveness of rating elicitation:} 
We observe that MOSTPOP produces the worst performance among all the baselines.
Since MOSTPOP does not employ rating elicitation, it provides a mere non-personalized recommendation to all new users.
As a result, the performance of MOSTPOP is even worse than RAN++.
With the rating elicitation, we can capture the preference of new users so that we can make personalized initial recommendations for them.

\vspace{1pt}
\noindent\textbf{Importance of the interactions between seed items:} 
We observe that the performance of the statistic-based methods (i.e., RAN++ and POP++) is significantly worse than the other approaches which consider the interactions between the seed items. 
With the consideration of inter-item interactions, we can find more representative seed itemset by avoiding the redundant selection.

\vspace{1pt}
\noindent\textbf{Importance of non-linear interactions and end-to-end training:}
We observe that \proposed outperforms the state-of-the-art competitor (i.e., RMVA) on all the datasets, especially on the CiteULike dataset.
Also, we find that the non-linear decoder is not always beneficial to improve the performance of the baselines
(RBMF vs. RBMF++ and RMVA vs. RMVA++).
As described in Section 3.1, the maximal-volume based methods only consider the linear interactions of the items both for finding the seed itemset and for reconstructing the user rating vector.
Although the decoder of RBMF++ and RMVA++ has a capability to capture the non-linearity, the performance is not always improved because the seed itemsets of them are not selected with the consideration of the non-linear interactions.
Unlike the existing approaches, \proposed optimizes the seed itemset and the non-linear decoder together through the end-to-end training, which enables the proposed framework to fully capture the complex structures of the CF information.  

\vspace{1pt}
\noindent\textbf{Deficiency of greedy selection:} 
In Figure 3, we observe that the performance of \proposed is more consistently improved on three datasets, compared to the maximal-volume approaches as the size of the seed itemset increases.
Because the greedy selection of the maximal-volume approaches cannot consider the interactions of the entire seed items at once, they may have limited capability of finding the most representative item combination.  
As a result, they cannot fully take advantage of the larger seed itemset.
However, \proposed can choose all the seed items at a time while considering non-linear interactions among them. 
This result again verifies the superiority of the proposed framework.

\begin{table}[t]
  \renewcommand{\arraystretch}{0.6}
  \caption{Effect of $\tau$ on P@20 on MovieLens 1M. The best result is in bold face.}
  \begin{tabular}{cccccc}
    \toprule
    \multirow{2}*{ }&\multirow{2}*{}& \multicolumn{4}{c}{$T_0$}\\
    \cmidrule{3-6}
    &&1&10&20&50\\
    \midrule
    \multirow{4}*{$T_E$}& 0.5 & 0.5324 & 0.5348 & 0.5336 & 0.5332\\
    & 0.1 & 0.5315 & \textbf{0.5396} & 0.5352 & 0.5281\\
    & 0.01 & 0.5105 & 0.5325 & 0.5374 & 0.5263 \\
   \cmidrule{2-6}
    & $T_0$ & 0.5158 & 0.4798 & 0.4502 & 0.4503 \\
    \bottomrule
  \end{tabular}
\end{table}

\subsection{Seed itemset Analysis}
For the qualitative comparison, we visualize the locations of the seed items in the latent space, where users' preferences on items are encoded, by using t-SNE \cite{tsne08}.
We train the latent vectors of users and items on MovieLens 1M with the state-of-the-art CF method for implicit feedback \cite{cml17}.
In Figure 4, we plot the latent vectors of users and items as blue dots on two-dimensional plane, and we mark the seed items with red translucent circles. 
We can clearly see that the seed items obtained by \proposed are more scattered around the plane than those of RMVA and RBMF.
Most seed items of RMVA and RBMF are densely located at the left-center of the plane. 
This result supports our argument that \proposed can select the better seed itemset compared to the maximal-volume approaches which may select redundant items in terms of non-linear dependency.

\subsection{Hyperparameter Analysis}
In this section, we analyze the effect of the hyperparameter $\tau$ with the result in Table 3. 
With the large $\tau$, the output of Gumbel-Softmax is smooth so that \proposed can explore the various combinations of items. 
With the annealed small $\tau$, the output of Gumbel-Softmax becomes one-hot so that \proposed can select the seed itemset concretely.
Note that the last row of Table 3 shows the result when $T_E = T_0$ (i.e., no temperature annealing). 
We observe that the temperature annealing is significantly beneficial to improve model performance.

\section{Conclusion}
We review the existing methods for the rating elicitation and point out two major flaws of them. 
Solving those two flaws, we propose \proposed, a novel deep learning framework that selects the seed itemset at a time with consideration of the non-linear interactions of the items.
To this end, \proposed first defines the categorical distributions to sample seed items, then trains both the categorical distributions and a non-linear decoder in the end-to-end manner.
After selecting the seed itemset, \proposed can provide personalized recommendation for a new user by using the trained decoder and the feedback on the seed itemset. 
With our extensive experiments, we show that \proposed outperforms all state-of-the-art methods in terms of two ranking metrics on four real-world datasets.
Also, we conduct a qualitative experiment to support our arguments about the superiority of \proposed.

\section*{Acknowledgements}
This research was supported by the MSIT(Ministry of Science and ICT), Korea, under the ICT Consilience Creative program(IITP-2019-2011-1-00783) supervised by the IITP(Institute for Information \& communications Technology Planning \& Evaluation).

\bibliographystyle{ACM-Reference-Format}
\bibliography{reference}

\end{document}